\documentclass[doublecol]{epl2} 
\usepackage{graphics}
\usepackage{graphicx}
\usepackage{amsmath}
\usepackage{amssymb}
\usepackage{amsthm}
\usepackage{epsfig}
\usepackage{dsfont}
\usepackage{xcolor} 
\usepackage[T1]{fontenc}
\usepackage{lmodern}
\title{Comment on "`Inconsistency of the conventional theory of superconductivity"' by Hirsch J.E.}
\author{Jacob Szeftel\inst{1} \and Nicolas Sandeau\inst{2} \and Michel Abou Ghantous\inst{3} \and Antoine Khater\inst{4}}
\institute{                  
  \inst{1} ENS Cachan, LPQM, 61 avenue du Pr\'esident Wilson, 94230 Cachan, France\\
	\inst{2} Aix Marseille Univ, CNRS, Centrale Marseille, Institut Fresnel, F-13013 Marseille, France\\
  \inst{3} American University of Technology, AUT Halat, Highway, Lebanon\\
	\inst{4} Universit\'e du Maine, UMR 6087 Laboratoire PEC, F-72000 Le Mans France}
\abstract{A rebuttal of arXiv : $1909.12786\quad (2019)$\cite{hir4} is presented below. To begin with, the main assumption, regarding the Meissner effect, which the whole argument relies upon, is disproved. Besides, the subsequent analysis misconstrues an original view of the Meissner effect by other authors\cite{sz2}. Finally, the discussion of the Joule effect turns out to violate the first law of thermodynamics.}
\begin{document}
\maketitle\par
		\section{An Incorrect Assumption}
Hirsch's rationale\cite{hir4} relies on an assumption, dating back to London's interpretation\cite{lon} of the Meissner effect, that the thermodynamical state, characterising a superconductor of type I, submitted to a time-dependent magnetic field $H(t)$, reaching eventually a constant value $H(t_f)$ for $t\geq t_f$, would depend \textit{only} on final temperature $T(t_f)$ and field $H(t_f)$ but would be conversely \textit{independent} from the \textit{transient} regime $(t<t_f)$, characterised by $\frac{dH}{dt}(t<t_f)\neq 0$ and (or) $\frac{dT}{dt}(t<t_f)\neq 0$. As shown elsewhere\cite{sz1,sz2}, such a claim, which entails furthermore that the skin-depth is independent from the frequency $\omega$ and equal to $\lambda_L/\sqrt{2}$ with $\lambda_L$ being London's length\cite{tin}, would be indeed true\cite{sz2}, if the electrical conductivity of the material were infinite. However, since the \textit{ac} conductivity was later measured to be \textit{finite}, albeit much \textit{larger} than the normal one, it was ascribed \textit{solely} to normal electrons\cite{tin}, while the superconducting ones were still believed to have infinite conductivity.\par
	Unfortunately, this mainstream view has been disproved\cite{sz3}, by showing on the basis of low-frequency susceptibility data\cite{max,str,ges,sar}, that the skin-depth was not equal to $\lambda_L/\sqrt{2}$ but was rather diverging like $1/\sqrt{\omega}$ for $\omega\rightarrow 0$, as seen in normal metals, and the conductivity, if ascribed to normal electrons, should be \textit{lower} than the normal one, in contradiction with experiment. In conclusion, contrary to a long-standing fallacy, the final $(t\geq t_f)$ state in the Meissner effect \textit{does} indeed depend\cite{sz3} on the whole \textit{transient} $(t< t_f)$ regime, due to irreversible consequences of \textit{finite ac} conductivity. Since Hirsch's main argument\cite{hir4} has been thereby rebutted, we could end our review at that point. But it is worth pursuing this subject, because the muddled discussion\cite{hir4} of the Meissner and Joule effects need clarification.
		\section{Meissner effect}
	Although Hirsch\cite{hir4} has long favored an interpretation of the Meissner effect, based on \textit{quantum pressure}\cite{hir3}, he suddenly embraces quite an unrelated explanation\cite{sz1,sz2,sz3}. In this novel view, the Meissner effect is ascribed to the susceptibility $\chi$, going from paramagnetic ($\chi_n>0$) in the normal ($T>T_c$) state to diamagnetic ($\chi_s<0$) in the superconducting ($T<T_c$) one ($T_c$ stands for the critical temperature). Despite $H$ remaining constant allover the cooling process, the magnetic induction $B$ is indeed altered at $T_c$ because of $\chi_s-\chi_n\neq 0\Rightarrow\frac{dB}{dt}\neq 0$, which gives rise, owing to the Faraday-Maxwell equation, to eddy currents flowing at the outer edge of the sample and screening $H$. Besides, due to the finite conductivity in the superconducting state\cite{sz2}, there is $\lambda_M>>\lambda_L$ with $\lambda_M$ being the penetration depth of $H$.\par
		Hirsch tries to apply\cite{hir4} this argument for $T<T_c$ by ascribing $\frac{d\chi}{dt}\neq 0$ to $\frac{d\lambda_L}{dT}\neq 0$ during the transient regime $\frac{dT}{dt}(t<t_f)\neq 0$. However, such a claim runs afoul\cite{sz2} at $\chi(T<T_c)\propto\left(\frac{\lambda_L}{\lambda_M}\right)^2$, which could only depend upon $T$ via the relaxation time\cite{ash} of the electron kinetic energy, $\tau$. Nevertheless, $\tau$ is very likely to be $T$ independent at such low temperature, for which it is limited by residual impurities.
		\section{Joule effect}
	Hirsch ascribes\cite{hir4} the whole Joule heat released during the transient regime to eddy currents, carried by normal electrons. However their contribution is negligible because the \textit{ac} conductivity of superconducting electrons can be larger than the normal one by $5$ five orders of magnitude\cite{ges}, which has been confirmed by analysing susceptibility data\cite{sz3}.\par
	The thermal balance, supposed to account\cite{hir4} for $T(t<t_f)$ in Eq.(12), turns out to violate the first law of thermodynamics\cite{lan} in two respects
\begin{itemize}
	\item
	the work, performed by the Faraday field and giving rise thereby to the eddy current, typical of the Meissner effect, has been overlooked. Strangely, a special emphasis is put on the sample being \textit{insulated} from the magnet producing $H$, which is tantamount to violating Newton's law, since the conduction electrons, conveying the eddy current, turn out to be accelerated without any force exerted on them;	
	\item
	only the heat, exchanged with an external reservoir, is considered\cite{hir4}, whereas that, released through the Joule effect, has been completely disregarded, although it plays a \textit{key} role in Hirsch's argument.	
\end{itemize}		
	 Therefore the issue of the Joule effect taking place in superconductors should be clarified. As a matter of fact, the Joule power $\dot{Q}_J=\frac{d{Q}_J}{dt}$, released  in a superconductor, has been shown elsewhere\cite{sz4} to comprise two contributions $\dot{Q}_1,\dot{Q}_2$ $\left(\Rightarrow\dot{Q}_J=\dot{Q}_1+\dot{Q}_2\right)$
\begin{itemize}
	\item	
	the usual one, warming up the sample, reads\cite{sz4} $\dot{Q}_1=\frac{j_s^2}{\sigma_s}$ with $j_s,\sigma_s$ standing for the density of the supercurrent and the \textit{finite} conductivity, associated with superconducting electrons. It is also equal to the work, performed by the Faraday field, apart from a tiny contribution\cite{sz5}, corresponding to the \textit{reversible} exchange between normal and superconducting electrons;
	\item
	the anomalous component equals\cite{sz4} $\dot{Q}_2=\frac{j_s^2}{\sigma_J}$ with $\sigma_J<0$ characterising the anomalous Joule effect, typical of superconductors, and causing the sample to \textit{cool} down. A simple experiment has been proposed\cite{sz4} to bring evidence for the anomalous Joule effect and to validate thereby a novel explanation of the persistent currents.
\end{itemize}		
	At last, it must be recalled that the specific heat of a superconductor depends\cite{sz5} upon the current flowing through it, because the current modifies the respective concentrations of normal and superconducting electrons.\par
		In summary, contrary to what is purported in ref.\cite{hir4}, the Meissner effect is found to depend upon the \textit{transient} regime\cite{sz2} and the Joule effect\cite{sz4} is seen to be \textit{consistent} with the theory of the normal to superconducting transition\cite{sz5} and the laws of thermodynamics\cite{lan}. 
		\acknowledgments
	One of us (J.S.) is indebted to P.W. Anderson for providing encouragement.

\end{document}